\documentstyle[aps]{revtex}
\begin{document}

\draft

\title{
Quasi-Local Conservation Equations in General Relativity}
\author{Jong Hyuk Yoon} 
\address{
Department of Physics, 
Konkuk University, Seoul 143-701, Korea\\
and \\
Asia Pacific Center for Theoretical Physics\\
Yoksam-dong 678-39, Kangnam-gu, Seoul 135-080, Korea\\
{\tt yoonjh@konkuk.ac.kr}}

\maketitle

\begin{abstract}
A set of exact quasi-local conservation equations is derived from 
the Einstein's equations using the first-order Kaluza-Klein 
formalism of general relativity in the (2,2)-splitting of 
4-dimensional spacetime. 
These equations are interpreted as quasi-local 
energy, momentum, and angular momentum conservation equations.
In the asymptotic region of asymptotically flat spacetimes, 
it is shown that the quasi-local energy and energy-flux integral 
reduce to the Bondi energy and energy-flux, respectively. 
In spherically symmetric spacetimes,
the quasi-local energy becomes the Misner-Sharp energy. 
Moreover, on the event horizon of a general dynamical black hole, 
the quasi-local energy conservation equation 
coincides with the conservation equation studied by Thorne {\it et al}. 
We discuss the remaining quasi-local conservation 
equations briefly. 
\end{abstract}


\begin{section}{Introduction and Kinematics}

In general relativity there have been many attempts to obtain 
quasi-local conservation 
equations\cite{bro-york93,hay94,nest91,sza94,berg92,yoon99c}.
One of the motivations of these efforts is that the quasi-local
conservation equations 
allow us to predict certain aspects of the future 
of a quasi-local region of a given spacetime 
without actually solving the Einstein's equations 
for that region.
Recall that in the Newtonian theory, the conservation of momentum
\begin{math}
\sum\vec{p}={\rm constant}
\end{math} \ 
immediately follows from Newton's third law,
which is no more than the consistency condition implementing 
Newton's second law. In general relativity, the consistency conditions 
for evolution are already incorporated 
into the Einstein's equations through the constraint equations,
from which {\it global} conservation equations were 
found. The purpose of this letter is to show that, from the 
Einstein's equations, one can find conservation equations of a stronger 
form, namely, {\it quasi-local} conservation equations. 
These equations, interpreted 
as quasi-local energy, momentum, 
and angular momentum conservation equations, are {\it exact} and  
{\it unique} in the sense that they are obtained 
by integrating the Einstein's ``constraint'' equations 
over a compact two-surface\cite{wini}.

Let us start from the following line
element
\cite{din-sta78,din-sma80,yoon92,yoon93a,yoon99a,yoon99b,new-unti62,new-tod80}
\begin{equation}
ds^2 = -2dudv - 2hdu^2 +{\rm e}^{\sigma} \rho_{ab}
 \left( dy^a + A_{+}^{\ a}du +A_{-}^{\ a} dv \right) 
\left( dy^b + A_{+}^{\ b}du +A_{-}^{\ b} dv 
\right),           \label{yoon}
\end{equation}
where $+,-$ stands for $u,v$, respectively. The geometry (\ref{yoon}) can 
be understood as follows. The hypersurface $u={\rm constant}$ 
is an out-going null hypersurface, 
and the hypersurface $v={\rm constant}$ is either timelike, null, 
or spacelike, depending on the sign of $2h$.
The intersection of two hypersurfaces $u,v={\rm constant}$ defines
a spacelike compact two-surface $N_{2}$, on which we introduce
the coordinates $y^{a}(a=2,3)$. The metric on $N_{2}$ 
is written as
\begin{equation}
\phi_{a b}={\rm e}^{\sigma}\rho_{ab},
\end{equation}
where $\rho_{ab}$ is the conformal two-metric satisfying 
the condition that 
\begin{equation}
{\rm det}\ \rho_{ab}=1,                        \label{det}
\end{equation}
and ${\rm e}^{\sigma}$ is the area element.
Notice that $v$ is the affine parameter of the null vector field 
\begin{equation}
{\partial \over \partial v} 
- A_{-}^{\ a}{\partial \over \partial y^{a}},  \label{nullv}
\end{equation}
which rules the out-going null hypersurface $u={\rm constant}$.
If we further assume that $A_{-}^{\ a}=0$, then the metric 
(\ref{yoon}) becomes identical to the Newman-Unti 
metric \cite{new-unti62}. 
In this letter, however, we shall retain the $A_{-}^{\ a}$ field,
since its presence will make the $N_{2}$-diffeomorphism invariant
{\it Yang-Mills} type gauge theory aspect of this  Kaluza-Klein 
formalism transparent. Let us mention that
the coordinates used in the above 
construction are not unique, which means that there are residual 
symmetries that preserve the metric (\ref{yoon}).
These residual symmetries consist of the diffeomorphisms of $N_{2}$, 
the reparametrization of $u$, and the shift of the origin of 
the affine parameter $v$ at each point of $N_{2}$, and were studied
in detail in \cite{new-tod80}. 

The spacetime integral $I_{0}$ of the scalar curvature of 
the metric (\ref{yoon}) can be written as
\begin{equation}
I_{0} =   \int \! \! du \, dv \, d^{2}y \, L_{0} 
+ {\rm surface}  \ {\rm integral},       \label{bareact}
\end{equation}
where
$L_{0}$ is given by\cite{yoon99c,yoon93a,yoon99a,yoon99b}
\begin{eqnarray}
& & L_{0} = -{1\over 2}{\rm e}^{2 \sigma}\rho_{a b}
  F_{+-}^{\ \ a}F_{+-}^{\ \ b}
  +{\rm e}^{\sigma} (D_{+}\sigma) (D_{-}\sigma)
  -{1\over 2}{\rm e}^{\sigma}\rho^{a b}\rho^{c d}
 (D_{+}\rho_{a c})(D_{-}\rho_{b d})
 +{\rm e}^{\sigma} R_2            \nonumber\\
& & -2{\rm e}^{\sigma}(D_{-}h)(D_{-}\sigma) 
- h {\rm e}^{\sigma}(D_{-}\sigma)^2
+{1\over 2}h  {\rm e}^{\sigma}\rho^{a b}\rho^{c d}
 (D_{-}\rho_{a c})(D_{-}\rho_{b d}).     \label{barelag}
\end{eqnarray}
Here $R_{2}$ is the scalar curvature of $N_{2}$, and 
we defined the diff$N_{2}$-covariant derivatives as follows,
\begin{eqnarray}
& &F_{+-}^{\ \ a}=\partial_{+} A_{-} ^ { \ a}-\partial_{-}
  A_{+} ^ { \ a} - [A_{+}, A_{-}]_{\rm L}^{a},  \label{field}\\
& &D_{\pm}\sigma = \partial_{\pm}\sigma
-[A_{\pm}, \sigma]_{\rm L},         \label{get}\\
& &D_{\pm}h= \partial_{\pm}h - [A_{\pm}, h]_{\rm L},   \label{eichid}\\
& &D_{\pm}\rho_{a b}=\partial_{\pm}\rho_{a b}
   - [A_{\pm}, \rho]_{{\rm L}a b},    \label{rhod}
\end{eqnarray}
where $\partial_{+}={\partial / \partial u}$, 
$\partial_{-}={\partial / \partial v}$, 
$\partial_{a}={\partial / \partial y^{a}}$, 
and $[A_{\pm}, \ast]_{\rm L}$ is the Lie derivative 
of $\ast$ along the vector field
$A_{\pm}:=A_{\pm}^{\ a}\partial_{a}$. 

In addition to the eight equations of motion that follow directly 
from the integral  $I_{0}$ by the variational principle, 
there are 
two supplementary equations associated with the partial gauge-fixing of
the general metric to the metric (\ref{yoon}). 
These equations, from which two quasi-local conservation equations
are shown to follow, are obtained by varying 
the Einstein-Hilbert action before the partial gauge-fixing 
condition is introduced. Here we present them without derivation,
and they are given by\cite{yoon99b} 
\begin{eqnarray}
& & ({\rm i}) \ 
-{\rm e}^{\sigma} D_{+}^{2}\sigma 
- {1\over 2}{\rm e}^{\sigma}(D_{+}\sigma)^{2}
-{\rm e}^{\sigma}(D_{-}h) (D_{+}\sigma) 
+{\rm e}^{\sigma}(D_{+}h)(D_{-}\sigma)     \nonumber\\
& & +2h {\rm e}^{\sigma}(D_{-}h)(D_{-}\sigma)  
+{\rm e}^{\sigma}F_{+-}^{\ \ a}\partial_{a}h  
-{1\over 4}{\rm e}^{\sigma}\rho^{a b}\rho^{c d} 
 (D_{+}\rho_{a c})(D_{+}\rho_{b d})    
+\partial_{a}\Big( \rho^{a b}\partial_{b}h \Big)   \nonumber\\
& & +h {\rm e}^{\sigma}
\Big\{R_{2} - (D_{+}\sigma) (D_{-}\sigma) 
 +{1\over 2}\rho^{a b}\rho^{c d} (D_{+}\rho_{a c})
    (D_{-}\rho_{b d}) 
+{1\over 2}{\rm e}^{\sigma}\rho_{a b}F_{+-}^{\ \ a}F_{+-}^{\ \ b}
 \Big\}         \nonumber\\
& & +h^{2}{\rm e}^{\sigma}\Big\{
(D_{-}\sigma)^{2}
-{1\over 2}\rho^{a b}\rho^{c d}
(D_{-}\rho_{a c}) (D_{-}\rho_{b d})\Big\}=0,  \label{gg} \\
& & ({\rm ii}) \ {\rm e}^{\sigma} D_{+}D_{-}\sigma 
+ {\rm e}^{\sigma} D_{-}D_{+}\sigma  
+ 2{\rm e}^{\sigma} (D_{+}\sigma)(D_{-}\sigma) %\nonumber\\
- 2{\rm e}^{\sigma}(D_{-}h)(D_{-}\sigma)   \nonumber\\
& & - {1\over 2}{\rm e}^{ 2 \sigma}\rho_{a b}
   F_{+-}^{\ \ a}F_{+-}^{\ \ b}
- {\rm e}^{\sigma} R_{2}     %\nonumber\\
- h {\rm e}^{\sigma} \Big\{
(D_{-}\sigma)^{2} 
-{1\over 2}\rho^{a b}\rho^{c d} 
 (D_{-}\rho_{a c})(D_{-}\rho_{b d})\Big\}=0. \label{ff}
\end{eqnarray}
\end{section}

\begin{section}{A Set of Quasi-local Conservation Equations}

Notice that the equations (\ref{gg}) and (\ref{ff}) are 
{\it first-order} in $D_{-}$ derivatives, so that they may 
be regarded as two ``constraint'' equations. 
Thus, in this formalism, 
the {\it natural} vector field that defines the evolution
is $\partial_{-}$.
Then the momenta 
\begin{math}
\pi_{I}=\{ \pi_{h},\pi_{\sigma}, \pi_{a}, \pi^{a b} \} 
\end{math} \
conjugate to the configuration variables 
\begin{math}
q^{I}=\{ h, \sigma,   A_{+} ^ { \ a}, \rho_{a b} \}
\end{math} \
are defined as 
\begin{equation}
\pi_{I}:={\partial L_{0}\over \partial (\partial_{-}{q}^{I}) }, \label{momenta}
\end{equation}
and are given by
\begin{eqnarray}
& &\pi_{h}=-2 {\rm e}^{\sigma}(D_{-}\sigma),   \label{pih}\\
& &\pi_{\sigma} = -2 {\rm e}^{\sigma} (D_{-}h)
       -2h {\rm e}^{\sigma} (D_{-}\sigma)
     + {\rm e}^{\sigma} (D_{+}\sigma),   \label{pisigma}  \\
& &\pi_{a}={\rm e}^{2 \sigma} \rho_{a b}F_{+-}^{\ \ b}, \label{pia} \\
& &\pi^{a b}= 
h{\rm e}^{\sigma} \rho^{a c}\rho^{b d}(D_{-}\rho_{c d})
-{1\over 2}{\rm e}^{\sigma} \rho^{a c}\rho^{b d}
(D_{+}\rho_{c d}).                            \label{first}
\end{eqnarray}
Notice that  $\pi^{a b}$ is traceless
\begin{equation}
\rho_{ab}\pi^{a b}=0
\end{equation}
due to the identities
\begin{equation}
\rho^{ab}D_{\pm}\rho_{ab}=0. 
\end{equation} 
The ``Hamiltonian'' function $H_{0}$ {\it defined} as 
\begin{equation}
H_{0}:=\pi_{I}\partial_{-}{q}^{I} - L_{0}
\end{equation}
is found to be
\begin{equation}
H_{0}= H -A_{-}^{\ a}C_{a} 
+ {\rm surface}\  {\rm  terms},           \label{ok}
\end{equation}                             
where $H$ and $C_{a}$ are given by
\begin{eqnarray}
& & H  =  -{1\over 2}{\rm e}^{-\sigma}\pi_{\sigma}\pi_{h} 
+ {1\over 4}h{\rm e}^{-\sigma}\pi_{h}^{2} 
-{1\over 2}{\rm e}^{-2\sigma}\rho^{a b}\pi_{a}\pi_{b} 
+{1\over 2h}{\rm e}^{-\sigma}
\rho_{a c}\rho_{b d}\pi^{a b}\pi^{c d}    \nonumber\\
& &
+{1\over 2}\pi_{h}(D_{+}\sigma)  
+{1\over 2h}\pi^{a b}(D_{+}\rho_{a b}) 
+{1\over 8h}{\rm e}^{\sigma}\rho^{a b} \rho^{c d}
(D_{+}\rho_{a c}) (D_{+}\rho_{b d}) 
-{\rm e}^{\sigma}R_{2},        \label{tilde}\\
& &C_{a}=\partial_{+}\pi_{a} 
-\partial_{b}(A_{+}^{\ b}\pi_{a})
-\pi_{b}\partial_{a}A_{+}^{\ b}
-\pi_{\sigma}\partial_{a}\sigma 
+ \partial_{a}\pi_{\sigma} - \pi_{h}\partial_{a}h
     - \pi^{b c}\partial_{a}\rho_{b c}   \nonumber\\
& &      + \partial_{b}( \pi^{b c}\rho_{a c})    
     + \partial_{c}( \pi^{b c}\rho_{a b})   
     - \partial_{a}( \pi^{b c}\rho_{b c}).       \label{ctilde}
\end{eqnarray}
In terms of these canonical variables $\{\pi_{I},{q}^{I}\}$,
the supplementary equations (\ref{gg}) and (\ref{ff}) 
can be written as
\begin{eqnarray}
& & ({\rm i'}) \
\pi^{a b}D_{+}\rho_{a b}   + \pi_{\sigma}D_{+}\sigma 
-h D_{+} \pi_{h}    
-\partial_{+}\Big( 
     h\, \pi_{h} + 2 {\rm e}^{\sigma}D_{+}\sigma \Big) \nonumber\\
& & +\partial_{a}\Big( 
    h\, \pi_{h}A_{+}^{ \ a}    
+ 2A_{+}^{ \ a}{\rm e}^{\sigma}D_{+}\sigma 
 + 2h {\rm e}^{-\sigma}\rho^{a b}\pi_{b}  
 +2\rho^{a b}\partial_{b}h \Big) =0,         \label{qenergy}\\
& & ({\rm ii'}) \ H - \partial_{+}\pi_{h} 
+ \partial_{a} \Big( 
A_{+}^{\ a}\pi_{h} 
+ {\rm e}^{-\sigma}\rho^{a b} 
\pi_{b} \Big)=0.                   \label{qmomentum}
\end{eqnarray}
Moreover, we have two more first-order equations
\begin{equation}
C_{a}=0,               \label{qangular}
\end{equation}
which follows trivially by varying $H_{0}$ with respect to 
$A_{-}^{\ a}$  
(or by varying the Einstein-Hilbert action\cite{yoon92,yoon99a},
which requires rather lengthy computations).
The equations (\ref{qenergy}), (\ref{qmomentum}), and
(\ref{qangular}) are the four Einstein's ``constraint'' equations
in the gauge (\ref{yoon}). 
Notice that the equations (\ref{qenergy}) and (\ref{qmomentum}) 
are {\it divergence}-type equations. 
If we contract the equation (\ref{qangular}) 
by an arbitrary function $\xi^{a}$ such that
\begin{equation}
\partial_{\pm} \xi^{a}=0,
\end{equation}
then the resulting equation is also a divergence-type equation, 
\begin{eqnarray}
& & 
\pi^{a b}{\pounds_{\xi}} \rho_{a b}
+\pi_{\sigma}{\pounds_{\xi}} \sigma
+\pi_{h}{\pounds_{\xi}} h
+\pi_{a}{\pounds_{\xi}} A_{+}^{\ a} 
-\partial_{+}( \xi^{a}\pi_{a} )   \nonumber\\
& & 
+\partial_{a}\Big(
-\xi^{a}  \pi_{\sigma} + 2 \pi^{a b} \xi^{c} \rho_{b c}
+A_{+}^{\ a} \xi^{b} \pi_{b} \Big)
=0,                                   \label{qangular2}
\end{eqnarray}
where $\pounds_{\xi}$ is the Lie derivative 
along $\xi:=\xi^{a}\partial_{a}$.

The integrals of these equations over a compact two-surface 
$N_{2}$ become, after a suitable normalization, 
\begin{eqnarray}
& & {\partial \over \partial u} U(u,v)
={1\over 16\pi}  \oint \! d^{2}y \, \Big(
\pi^{a b}D_{+}\rho_{a b} +\pi_{\sigma}D_{+}\sigma  
-h D_{+} \pi_{h}
   \Big),              \label{enflux} \\
& &  {\partial \over \partial u} P(u,v)
={1 \over 16\pi} \oint \! d^{2}y \, H,  \label{momflux}  \\
& & {\partial \over \partial u} L(u,v;\xi)
={1 \over 16\pi}\oint \! d^{2}y \, \Big( 
\pi^{a b}\pounds_{\xi} \rho_{a b}  
+\pi_{\sigma}\pounds_{\xi} \sigma 
-h \pounds_{\xi} \pi_{h} 
-A_{+}^{\ a}\pounds_{\xi}\pi_{a} \Big),  \label{angflux} 
\end{eqnarray}
where  $U(u,v)$, $P(u,v)$, and $L(u,v;\xi)$ are defined as
\begin{eqnarray}
& & U(u,v):= {1 \over 16\pi}\oint d^{2}y \,   \Big( 
h\, \pi_{h} + 2 {\rm e}^{\sigma} D_{+}\sigma \Big) 
+ U^{0},               \label{enint}\\
& &
P(u,v):={1 \over 16\pi} \oint \! d^{2}y \, ( \pi_{h} )
+ P^{0},                \label{momint}\\
& &
L(u,v;\xi):={1 \over 16\pi} \oint \! d^{2}y \, 
(\xi^{a}\pi_{a}) + L^{0}.       \label{angint}
\end{eqnarray}
Here $H$ is the Hamiltonian function given by (\ref{tilde}),
and $U^{0}$, $P^{0}$, and $L^{0}$ are undetermined 
subtraction terms, which however must be $u$-independent,
\begin{equation}
{\partial U^{0}\over \partial u} 
={\partial P^{0}\over \partial u}
={\partial L^{0}\over \partial u} =0,    \label{subtract}
\end{equation}
in order to satisfy the equations (\ref{enflux}), (\ref{momflux}), 
and (\ref{angflux}), respectively. 

One can write the r.h.s. of the equation (\ref{enflux}) 
in a more symmetric and suggestive form as follows.
Let us contract the equation (\ref{qangular}) with 
$A_{+}^{\ a}$
and integrate the resulting equation over $N_{2}$ 
to obtain the following equation,
\begin{equation}
\oint \! d^{2}y \, \Big( 
A_{+}^{\ a} \partial_{+}\pi_{a} \Big)  
=\oint \! d^{2}y \, \Big( 
\pi^{a b}{\pounds_{A_{+}}} \rho_{a b}  
+\pi_{\sigma}{\pounds_{A_{+}}} \sigma 
-h {\pounds_{A_{+}}} \pi_{h} \Big).  \label{xxx} 
\end{equation}
If we use the definitions of diff$N_{2}$-covariant 
derivatives $D_{\pm}$ and 
the equation (\ref{xxx}), then the equation (\ref{enflux}) 
can be written as
\begin{equation}
{\partial \over \partial u} U(u,v)
={1\over 16\pi}  \oint \! d^{2}y \, \Big(
\pi^{a b}  \partial_{+} \rho_{a b} 
+\pi_{\sigma}\partial_{+}\sigma  - h \partial_{+} \pi_{h}
-A_{+}^{\ a} \partial_{+}\pi_{a} \Big).   \label{enflux1}
\end{equation}
Notice that the r.h.s. of the conservation equations (\ref{angflux}) 
and (\ref{enflux1}) match exactly, if we interchange the 
derivatives
\begin{equation}
{\pounds_{\xi}}  \longleftrightarrow \partial_{+} \label{match}
\end{equation}
in the two-surface integrals.

In a region of a spacetime where
\begin{math}
\partial / \partial u
\end{math} \ 
is timelike, these quasi-local conservation equations 
relate the instantaneous 
rates of changes of two-surface integrals at a given $u$-time
to the associated net flux integrals,
and they form a ``complete'' set of quasi-local conservation equations
since they follow directly from the four Einstein's ``constraint'' 
equations. Let us remark that,
unlike the Tamburino-Winicour's quasi-local conservation 
equations\cite{wini} which are ``weak'' conservation equations 
since they assumed in the derivation the Ricci flat conditions
(i.e. the full Einstein's equations), 
our quasi-local conservation equations are ``strong'' 
conservation equations since we used the four Einstein's ``constraint'' 
equations only. 

\end{section}

\begin{section}{Quasi-Local Energy and Energy-Flux Integral}

The equations (\ref{enflux}), (\ref{momflux}), and 
(\ref{angflux})
are in fact quasi-local energy, momentum, and angular momentum 
conservation equations, respectively\cite{wini,yoon00a}.
In this section we shall focus on the quasi-local energy 
conservation equation (\ref{enflux}), and defer discussions of 
other conservation equations to the section \ref{discuss}. 
In order to define a quasi-local energy associated with 
a given two-surface $N_{2}$, 
we have to introduce a subtraction term  $U^{0}$ referring 
to that region only.
In general the subtraction terms for a quasi-local energy are
not unique, and the ``right'' subtraction term may not even exist 
at all in a generic situation. One natural criterion is 
that the subtraction term must be chosen such that the 
quasi-local energy reproduces ``standard'' values 
in limiting cases. One possible candidate is 
\begin{equation}
U^{0}= \sqrt{{\cal A}\over 16\pi},      \label{ezero}
\end{equation}
where ${\cal A}$ is the area of $N_{2}$. However, 
this subtraction term introduces
a restriction on the admissible two-surfaces $N_{2}$ or, 
equivalently, on the vector field 
\begin{math}
{\partial / \partial u},
\end{math}
due to the condition
\begin{equation}
{\partial {\cal A} \over \partial u}=0         \label{azero}
\end{equation}
that follows from (\ref{subtract}).
But this is just the condition that the Bondi time function 
satisfies at the null infinity.
Therefore the time function $u$ satisfying the condition (\ref{azero}) 
may be regarded as a quasi-local, finite distance analog 
of the standard Bondi time function. The condition (\ref{azero}) 
can be interpreted as follows: In order to evaluate and 
compare quasi-local energies of a given two-surface $N_{2}$ 
at two different times in a physically meaningful way, 
there must exist certain requirements on $N_{2}$,
and the minimum requirement is that the area of the two-surface 
$N_{2}$  is independent of $u$.

The quasi-local energy integral for a two surface with 
the area ${\cal A}$ is then given by
\begin{equation}
U(u,v)= {1 \over 16\pi}\oint d^{2}y \,   \Big( 
h\, \pi_{h} + 2 {\rm e}^{\sigma} D_{+}\sigma \Big) 
+ \sqrt{{\cal A}\over 16\pi},             \label{moden}
\end{equation}
which in general is not positive-definite, and does not 
possess monotonicity either. The lack of these properties seems to be
related to the fact that gravitational binding energy is 
always negative. 
However, as we shall see shortly, the quasi-local energy function 
at the null infinity reduces to the Bondi mass, which is 
positive-definite
and decreases monotonically as the time $u$ increases.

\vspace{.7cm}
%\begin{subsection}{The Bondi mass-loss relation}

\noindent
{\it III-1. The Bondi mass-loss formula}\\
In asymptotically flat spacetimes, the metric 
becomes
\begin{equation}
ds^2 \longrightarrow 
-2dudv-\Big( 1-{2m\over v}\Big)\, du^2
 + v^{2} d\Omega^{2}        \label{sol}
\end{equation}
as $v \rightarrow \infty$. Thus 
\begin{math}
\partial / \partial u
\end{math} \ 
is asymptotic to the timelike Killing vector field. 
In general, the energy-flux across a two-surface is given by
the energy-momentum tensor $T_{0i}$, which is of the form
\begin{equation}
T_{0i}\sim \pi_{\phi} \partial_{i}\phi.
\end{equation}
The integrand of the r.h.s. of (\ref{enflux1}) is of this
form, and we therefore expect that it represents 
the energy-flux carried by gravitational radiation crossing $N_{2}$. 
Then the l.h.s. of (\ref{enflux1}) should be 
the instantaneous rate of change in the gravitational energy 
of the region enclosed by $N_{2}$. 
The energy-flux integral in general does not have a definite sign, 
since it includes the energy-flux carried by the in-coming 
as well as the out-going gravitational radiation. 
But in the asymptotically flat region, 
the energy-flux integral turns out to be negative-definite, 
representing the physical situation that 
there is no in-coming flux coming from the infinity.

Let us now show that the equation (\ref{enflux}) 
reduces to the Bondi mass-loss formula\cite{new-tod80}
in the asymptotic region of asymptotically flat spacetimes.
The asymptotic fall-off rates of the metric coefficients 
in the asymptotic Bondi coordinates 
are given by\cite{wald,falloff} 
\begin{eqnarray}
& &{\rm e}^{\sigma}=v^{2}   ({\rm sin}\vartheta) \Big\{ 
1 +O({1 \over v^{2}})  \Big\},           \label{fallsig}\\
& &\rho_{\vartheta \vartheta}
=\Big( {1\over {\rm sin}\vartheta} \Big) \Big\{ 
1 + {\alpha \over v}
+ O({1 \over v^{2}}) \Big\},       \label{fallrho1} \\  
& &\rho_{\varphi \varphi}=({\rm sin}\vartheta)  \Big\{ 
    1 + {\beta \over v} 
+ O({1 \over v^{2}}) \Big\},    \label{fallrho2} \\
& &\rho_{\vartheta \varphi}
={\gamma \over v}+ O({1 \over  v^{2}}),   \label{fallrho3}\\
& &2h=1-{2m \over v} 
 + O({1 \over  v^{2}}),         \label{fallh}\\
& &A_{+}^{\ a}=O({1 \over v}), \\ %\hspace{0.5cm} 
& &A_{-}^{\ a}=O({1 \over v^{2}}),       \label{falla}
\end{eqnarray}
where the expansion coefficients $\alpha, \beta, \gamma$, and $m$ 
are functions of $(u,\vartheta,\varphi)$. 
Then the total energy at the null infinity coincides with 
the Bondi energy $U_{\rm B}(u)$,
\begin{equation}
\lim_{v \rightarrow \infty}U(u,v) 
={1 \over 4\pi}\oint_{S_{2}} \!\!\! 
d \Omega \, \, m (u, \vartheta,\varphi)= U_{\rm B}(u),
\end{equation}
where $m(u,\vartheta,\varphi)$ is the mass aspect 
of the asymptotically flat radiating spacetime.
One can easily show that the equation (\ref{enflux}) becomes 
\begin{equation}
{d \over d u} U_{\rm B}(u) 
=-{1\over 32\pi}
  \oint \! d\Omega \, v^{2}
  \rho^{a b}\rho^{c d}
(\partial_{+}\rho_{a c})(\partial_{+}\rho_{b d}) 
\leq 0,                            \label{lindo}
\end{equation} 
which is just the Bondi mass-loss formula. Notice that
the negative-definite energy-flux is a bilinear of 
the traceless {\it current} $j^{a}_{ \ b}$ defined as
\begin{equation}
j^{a}_{ \ b}:=\rho^{a c}\partial_{+}\rho_{ b c} \hspace{.5cm}
(j^{a}_{ \ a}=0), 
\end{equation}
representing the shear degrees of freedom 
of gravitational radiation.

%\end{subsection}

\vspace{.7cm}
%\begin{subsection}{The Misner-Sharp energy}

\noindent
{\it III-2. The Misner-Sharp energy}\\
Let us now consider a spherical ball
of radius $v$ filled with a perfect fluid with energy density 
$\rho(v)$. 
The Einstein's equations are modified by the presence of 
the fluid, and the solution is given by 
\begin{equation}
ds^2 = -2dudv - 2h(v)du^2 + v^{2} 
    d \Omega ^{2},             \label{fluid}
\end{equation}
where 
\begin{eqnarray}
& & 2h(v)=1-{2 m(v)\over v},  \label{twoem}\\ %\hspace{0.5cm}
& & m(v)=4\pi \! \! \int_{0}^{v} \! \! 
d v' v'^{2}\, \rho(v').      \label{twoei}
\end{eqnarray}
If we choose the subtraction term such that
\begin{equation}
U^{0}(v)=\sqrt{{\cal A}\over 16\pi}={v\over 2},  \label{subtr2}
\end{equation}
then $U(v)$ becomes the Misner-Sharp energy 
$m(v)$\cite{mis-sha64,hay96}
\begin{equation}
U(v) =  m(v).       \label{asycorrecto}
\end{equation}

%\end{subsection}

\vspace{.7cm}
%\begin{subsection}{Black Hole}

\noindent
{\it III-3. Black holes}\\
One might be also interested in applying this formalism 
to black holes, 
and try to obtain quasi-local energy of the event horizon 
and quasi-local energy-flux incident on the horizon.
For this problem, it is appropriate to choose a coordinate system
adapted to the {\it in}-going null geodesics.
In a spacetime where the metric is given by 
\begin{equation}
ds^2 = + 2dudv - 2hdu^2 +{\rm e}^{\sigma} \rho_{ab}
 \left( dy^a + A_{+}^{\ a}du +A_{-}^{\ a} dv \right)  
\left( dy^b + A_{+}^{\ b}du +A_{-}^{\ b} dv 
\right),                  \label{ingoing}
\end{equation}
the vector field 
\begin{equation}
{\partial \over \partial v} 
- A_{-}^{\ a}{\partial \over \partial y^{a}}  \label{minus}
\end{equation}
is an in-going null vector field generating the null 
hypersurface $u={\rm constant}$, and the vector field 
\begin{equation}
{\partial \over \partial u} 
- A_{+}^{\ a}{\partial \over \partial y^{a}} \label{plus}
\end{equation}
whose norm is $-2h$ is 
either timelike, null, or spacelike, depending on the
sign of $2h$. Thus, in a region of a spacetime where $2h=0$, 
it becomes an out-going null vector field. 
Therefore, on the event horizon H
generated by the out-going null vector fields,
we must have $2h=0$.  If we repeat the previous analysis 
in the in-going null coordinate system (\ref{ingoing}), 
we obtain another set of quasi-local conservation equations.
On the event horizon H, the quasi-local energy conservation 
equation becomes
\begin{eqnarray}
& & {\partial U_{\rm H}\over \partial u}
={1\over 16\pi}  \oint_{\rm H} \! d^{2}y \, \Big\{
{1\over 2} {\rm e}^{\sigma} \rho^{a b}\rho^{c d}
(D_{+}\rho_{a c}) (D_{+}\rho_{b d})  
-{\rm e}^{\sigma} (D_{+}\sigma)^2  
  -2 {\rm e}^{\sigma} \kappa D_{+}\sigma
   \Big\},              \label{enfluxa} \\
& & U_{\rm H}:= -{1 \over 8\pi}
\oint_{\rm H} d^{2}y \,    
({\rm e}^{\sigma} D_{+}\sigma ) 
+ U_{\rm H}^{0},               \label{eninta}
\end{eqnarray}
where $\kappa:=D_{-}h|_{\rm H}$ is the {\it surface gravity} on
H. This equation is identical to the quasi-local energy 
conservation equation on the {\it stretched} 
horizon, which was studied in detail in \cite{membrane}. 
Notice that when the subtraction term 
$U_{\rm H}^{0}$ is chosen zero, the quasi-local energy $U_{\rm H}$
is non-positive since the area of the event horizon always increases,
\begin{equation}
D_{+}\sigma|_{\rm H} \geq 0.
\end{equation}
However, when the black hole no longer expands so that
$D_{+}\sigma|_{\rm H}=0$, then
$U_{\rm H}$ becomes zero. For instance, 
for a Schwarzschild or Kerr black hole, we have 
$U_{\rm H}=0$\cite{membrane,lbk,katz,def}. This counter-intuitive 
aspect is a manifestation of the well-known teleological nature of 
the event horizon. That is, when the event horizon H evolves, 
its quasi-local energy must be negative so as to cancel 
out the positive in-flux of energy carried by subsequently 
in-falling matter or gravitational radiation, 
leaving  $U_{\rm H}=0$ when the black hole 
reaches the final stationary state. 
Details of this derivation  and discussions of the remaining 
quasi-local conservation equations on the event horizon 
will be presented elsewhere\cite{yoon00a}.

%\end{subsection}
\end{section}

\begin{section}{Discussions}
\label{discuss}

In summary we derived a set of four quasi-local 
conservation equations from the Einstein's ``constraint'' equations. 
In particular, we showed that  
one of the quasi-local conservation equations reproduces
the Bondi mass-loss relation in the asymptotic region of 
asymptotically flat spacetimes, 
and that the quasi-local energy coincides with
the Misner-Sharp energy for spherically symmetric fluids 
in a finite region. We also applied the quasi-local energy conservation
equation to the horizon of a general dynamic black hole, and 
found that it reduces to the quasi-local conservation equation 
of Thorne {\it et al}.
It must be stressed that our quasi-local conservation equations 
are exact and unique, in that 
they were obtained directly from the Einstein's ``constraint'' 
equations through the first-order canonical formalism. 

It seems appropriate to mention that the equation (\ref{momflux}) 
has a similar structure to the integrated Navier-Stokes equation
for a viscous fluid\cite{lan-lif89},
\begin{equation}
{\partial P_{i} \over \partial u} 
= - \oint \!d S^{k} \Big(   
p\delta_{ik} +\rho v_{i}v_{k} -\sigma'_{ik}  \Big), \label{navier}
\end{equation}
where $ P_{i}$ and $\sigma'_{ik}$ are the total momentum
and the viscous term, 
\begin{eqnarray}
& & P_{i} = \int \! d V \, (\rho v_{i}),   \label{euler1}\\
& & \sigma'_{ik} =\eta\Big(  
{\partial v_{i} \over \partial x^{k}} 
+{\partial v_{k} \over \partial x^{i}} 
-{2\over 3}\delta_{ik}{\partial v_{l} \over \partial x^{l}} \Big)
+\zeta\delta_{ik}{\partial v_{l} \over \partial x^{l}}, \label{euler2}
\end{eqnarray}
and $\eta$ and $\zeta$ are the coefficients of 
shear and bulk viscosity, respectively. 
This equation tells us that the rate of the net momentum change 
of a fluid within a given volume is determined by
the net momentum-flux across the 
two-surface enclosing the volume. Notice that
the Hamiltonian function $H$ in (\ref{tilde}) 
is at most quadratic in the conjugate momenta $\pi_{I}$, 
and assumes a form of momentum-flux of a viscous fluid. 
From this point of view, terms quadratic in $\pi_{I}$ are 
responsible for direct momentum transfer, 
terms linear in $\pi_{I}$ may be regarded as viscosity terms, 
and terms independent of $\pi_{I}$ as pressure terms. 
This observation allows us to interpret the Hamiltonian function
$H$ as the gravitational momentum-flux and 
the two-surface integral
\begin{equation}
{1 \over 16\pi} \oint \! d^{2}y \, ( \pi_{h} ) 
\end{equation}
as the quasi-local gravitational momentum associated with $N_{2}$. 

The equation (\ref{angflux}) is a quasi-local angular momentum 
conservation equation, since the r.h.s. assumes the canonical form
\begin{equation}
\pi_{I}\pounds_{\xi}q^{I},
\end{equation}
representing the angular momentum-flux associated with 
the vector field $\xi=\xi^{a}\partial_{a}$. The proposed quasi-local 
momentum and angular momentum conservation equations will be
analyzed in detail in a separate paper\cite{yoon00a}.

A final remark concerns with quantum gravity. By replacing 
the conjugate momenta $\pi_{I}$ as 
\begin{equation}
\pi_{I} \rightarrow -i {\delta \over \delta q^{I}}  \label{quantum}
\end{equation}
in (\ref{enint}), (\ref{momint}), and (\ref{angint}), 
one can obtain a set of functional Schr{\"o}dinger equations. 
For instance, from the equation (\ref{enint}), one obtains the 
following equation,
\begin{equation}
{1 \over 16\pi}\oint d^{2}y \,   \Big\{
-i h {\delta \Psi \over \delta h} 
+ 2 {\rm e}^{\sigma} (D_{+}\sigma) \Psi \Big\}
=i {\partial \Psi \over \partial u}, 
\end{equation}
where the subtraction term $U^{0}$ was chosen zero. 
These functional Schr{\"o}dinger equations seem worth 
exploring in situations where quantum gravity effects are expected 
to be dominant, for example, near black holes.

\end{section}

\bigskip\noindent       
\centerline{\bf Acknowledgments}\\

It is a great pleasure to thank Dr. Gungwon Kang and Prof. 
Joohan Lee for several 
enlightening discussions, and Prof. Y. Choquet-Bruhat 
for suggesting 
the author to examine black holes using this (2,2)-formalism. 
This work was supported by Konkuk University research grant in 1999. 

\nopagebreak


\begin{references}
\bibitem{falloff}{The $O(1/v)$-term was removed from the expansion 
(\ref{fallsig}) by a suitable choice of the origin 
of the affine parameter $v$.}
\end{references}
\end{document}